\documentclass[%
 reprint,
superscriptaddress,
 amsmath,amssymb,
 aps,
prb,
]{revtex4-2}

\usepackage{graphicx}
\usepackage{dcolumn}
\usepackage{bm}

\begin{document}

\title{Model-independent determination of the gap function of nearly localized superconductors}

\author{Du\v{s}an Kavick\'{y}}
\affiliation{Department of Experimental Physics, Comenius University,
  Mlynsk\'{a} Dolina F2, 842 48 Bratislava, Slovakia}
\author{Franti\v{s}ek Herman}
\affiliation{Department of Experimental Physics, Comenius University,
  Mlynsk\'{a} Dolina F2, 842 48 Bratislava, Slovakia}
\affiliation{Institute for Theoretical Physics, ETH Z\"{u}rich, CH-8093, Switzerland}
\author{Richard Hlubina}
\affiliation{Department of Experimental Physics, Comenius University,
  Mlynsk\'{a} Dolina F2, 842 48 Bratislava, Slovakia}
  
\date{\today}

\begin{abstract}
The gap function $\Delta(\omega)$ carries essential information on both, the pairing glue as well as the pair-breaking processes in a superconductor. Unfortunately, in nearly localized superconductors with a non-constant density of states in the normal state, the standard procedure for extraction of $\Delta(\omega)$ cannot be applied. Here, we introduce a model-independent method that makes it possible to extract $\Delta(\omega)$ also in this case. The feasibility of the procedure is demonstrated on the tunneling data for the disordered thin films of TiN. We find an unconventional feature of $\Delta(\omega)$ which suggests that the electrons in TiN are coupled to a very soft pair-breaking mode.
\end{abstract}	
\maketitle

\section{Introduction}
\label{sec:introduction}
A prototypical example of a quantum phase transition is provided by the quantum breakdown of superconductivity (QBS) which may occur when a suitable external parameter is changed \cite{Sacepe20}. In the special case when the QBS is due to increasing disorder, which is the subject of this paper, there exist at least three mechanisms which can drive the transition: (i) Within the so-called fermionic scenario \cite{Anderson83,Finkelstein94}, increasing disorder leads to an increase of the repulsive part of the effective electron-electron interaction, thereby weakening the tendency towards superconductivity. Moreover, interaction-induced quantum corrections which lead to a suppressed density of states in the normal state \cite{Altshuler79} further weaken the BCS instability. (ii) Whatever causes the initial suppression of the superconducting critical temperature $T_c$, at some point the phase stiffness of a dirty superconductor necessarily becomes small and the phase fluctuations may destroy superconductivity. This mechanism is emphasized within the so-called bosonic scenario of the QBS \cite{Doniach81,Fisher90}. (iii) Finally, the single-particle wavefunctions become strongly inhomogeneous before undergoing Anderson localization. Theories which concentrate on this aspect of the QBS are known as the "emergent-granularity" scenario \cite{Ma85,Feigelman10}.

When the QBS occurs in homogeneously disordered systems, all three mechanisms are intertwined: (i) The relevance of the fermionic scenario is documented most convincingly by the recent observation of quantum corrections to optical conductivity up to at least $\sim$4~eV \cite{Neilinger19}. (ii) The temperature dependence of the resistivity indicates that phase fluctuations definitely can not be neglected in such systems \cite{Postolova20,Carbillet20}. (iii) The measured spatial modulations of the tunneling density of states strongly suggest that also electronic inhomogeneity in structurally effectively homogeneous systems does play a role \cite{Postolova20,Carbillet20}. Unfortunately, the combined effect of changing interactions, phase fluctuations, and localization is presently not understood \cite{Sacepe20}.

The symptoms of the various mechanisms driving the QBS can be further looked for, e.g., by studying the scaling of $T_c$ with the distance from the critical point, the magnitude of the gap-to-$T_c$ ratio, and the value of the critical resistivity \cite{Sacepe20}. However, a truly comprehensive picture of the transition would be provided by the knowledge of the gap function $\Delta(\omega)$ which is known to carry information not only on the pairing glue \cite{McMillan65}, but, as emphasized recently, also on the pair-breaking processes \cite{Bzdusek15,Kavicky20} which occur in a superconductor. Equally important is also information on the spatial dependence of the gap function $\Delta(\omega)$.

In case of conventional superconductors the gap function $\Delta(\omega)$ can be routinely extracted from the tunneling density of states in the superconducting state, $N_s(\omega)$ \cite{McMillan65}. For nearly localized superconductors, the tunneling data are in fact available from a series of recent papers \cite{Postolova20,Carbillet20,Zemlicka20}, but all of them find that the normal-state tunneling conductivity is not constant, presumably due to interaction-induced corrections \cite{Altshuler79}. 

It should be noted that the energy dependence of the tunneling conductivity might be influenced also by the tunneling process itself and in that case the extraction of the density of states from conductivity data may be problematic. One example of such extrinsic effects is provided by the energy dependence of the tunneling matrix elements \cite{Fischer07}. Alternatively, the tunneling process may be assisted by the environment \cite{Ingold92}. In principle, however, such effects can be excluded, e.g., if the tunneling spectra do not depend on the tip-to-sample distance. Therefore we do not take them into account in the present work and, at low temperatures, we do not distinguish between the tunneling conductivity and the density of states.

Unfortunately, if the normal-state density of states $N_n(\omega)$ is not constant, then the standard procedure for extraction of the pairing glue \cite{McMillan65} is not applicable. The goal of this paper is to demonstrate that, provided also $N_n(\omega)$ is known from experiment, the superconducting gap function $\Delta(\omega)$ of isotropic s-wave superconductors can nevertheless be determined, and the proposed procedure does not rely on any particular microscopic model. 

The outline of this paper is as follows. In Sec.~\ref{sec:elimination} we show how the nontrivial energy dependence of $N_n(\omega)$ can be eliminated from the measured superconducting density of states $N_s(\omega)$. The output of such elimination procedure is what we call the dos-function $n(\omega)$, which can be thought of as the density of states of a hypothetical superconductor which is otherwise identical to the studied one, but its normal-state density of states is constant.

Once the dos-function $n(\omega)$ is known, the next problem is how to extract the gap function $\Delta(\omega)$ from it. This task is similar to the McMillan-Rowell inversion \cite{McMillan65}, but the crucial difference has to do with experimental indications that the classical Eliashberg theory does not apply to very dirty superconductors \cite{Dynes86}. Therefore the inversion procedure should not make use of the concept of the pairing glue as in the standard approach \cite{McMillan65}, and our solution to this problem is presented in Sec.~\ref{sec:delta}. 

The feasibility of both steps of the procedure is demonstrated by their application to the tunneling data for homogeneously disordered thin films of TiN, which are taken from \cite{Postolova20}. 

In Sec.~\ref{sec:conclusions} we point out the limitations of our analysis and we suggest how future experiments could remove the ambiguities of the current results.

\section{Elimination of the nontrivial normal-state effects}
\label{sec:elimination}
It is well known that, in dirty systems at not too strong coupling, interaction-induced corrections to the normal-state density of states are present only if the electron self-energy depends on the energy $\varepsilon$ of the one-particle eigenstates  \cite{Rickayzen80,McMillan81,note_mott}. On the other hand, in order to take into account the retarded phonon-mediated pairing interactions, it is necessary to include also the frequency $\omega$ dependence of the self-energy. Thus, in order to allow for both phenomena, a generalized version of the Eliashberg theory with an $\varepsilon$- and $\omega$-dependent electron self-energy is needed for nearly localized superconductors. Such a theory has in fact been developed long ago \cite{Belitz89} and a model calculation of the superconducting density of states $N_s(\omega)$ in that formalism became available recently \cite{Rabatin18}. 

The discussion of the present paper starts from the results obtained in \cite{Rabatin18}. There it was assumed that the effective electron-electron interaction consists of two terms. The first term, due to poorly screened Coulomb interactions, is described by an $\varepsilon$-dependent pseudopotential $\mu(\varepsilon)$. The second term, characterized by an $\omega$-dependent coupling function $g(\omega)$, describes the standard phonon-mediated electron-electron interaction. Starting from this assumption, one could show that the electron self-energy in the superconducting state is described by four functions: the two standard frequency-dependent functions $Z(\omega)$ and $\Delta(\omega)$ known from the Eliashberg theory, plus two other functions $\chi(\varepsilon)$ and $\psi(\varepsilon)$, which describe the effect of the Coulomb pseudopotential \cite{Rabatin18}. The functions $Z(\omega)$ and $\chi(\varepsilon)$ quantify the diagonal part of the self-energy, whereas $\Delta(\omega)$ and $\psi(\varepsilon)$ describe its off-diagonal (i.e., superconducting) part. It is important to point out that this result for the self-energy is fairly general and does not depend on the particular form of the functions $\mu(\varepsilon)$ and $g(\omega)$. 

In \cite{Rabatin18}, a particular but reasonable model of the coupling functions $\mu(\varepsilon)$ and $g(\omega)$ has been studied. Two results of that study are of direct relevance to the present work. First, it was found that the self-energy $\chi(\varepsilon)$, which describes the $\mu(\varepsilon)$-induced change of the electron dispersion, does not change appreciably between the normal and the superconducting state. Second, the $\varepsilon$-dependent part of the anomalous self-energy $\psi(\varepsilon)$ was found to be small with respect to the $\omega$-dependent part of the anomalous self-energy. 

This leads us to assume that the weakly screened Coulomb interactions can be always described by a single temperature-independent self-energy $\chi(\varepsilon)$ - in other words, the self-energy $\psi(\varepsilon)$ can be neglected. It is obvious that, in presence of a finite self-energy $\chi(\varepsilon)$, the density of $\varepsilon$-levels changes to an $\varepsilon$-dependent function $N_0(\varepsilon)$, and one should expect that this function exhibits a minimum close to the Fermi level \cite{Altshuler79}.

At this point we depart from the discussion in \cite{Rabatin18} and, instead of applying the formalism to a specific model of the superconductor, we take a model-independent approach. In particular, we assume that the density of $\varepsilon$-levels is described by an arbitrary function $N_0(\varepsilon)$ with a minimum close to the Fermi level. We emphasize that we do not need to specify the functional form of $N_0(\varepsilon)$ and we do not even have to require that it is particle-hole symmetric. Then, in presence of a finite electron-phonon coupling, the observable density of states is by definition described by
\begin{equation}
N_i(\omega)=\int d\varepsilon N_0(\varepsilon) A_i(\varepsilon,\omega),
\label{eq:dos_definition}
\end{equation}
where $A_i(\varepsilon,\omega)$ is the spectral function of an electron in a~one-particle eigenstate with energy $\varepsilon$. The index $i=n,s$ discriminates between the normal and superconducting states. 

In order to proceed, in what follows we will evaluate the spectral functions $A_i(\varepsilon,\omega)$ within the Eliashberg theory. For future convenience it is useful to parametrize the two complex Eliashberg functions, the wavefunction renormalization $Z(\omega)$ and the gap function $\Delta(\omega)$, by four real functions of frequency $\widetilde{\omega}(\omega)$, $\widetilde{\gamma}(\omega)$, $\widetilde{\Omega}(\omega)$, and $\widetilde{\Gamma}(\omega)$, defined as follows:
\begin{eqnarray*}
Z(\omega) \omega&=&\widetilde{\omega}+i\widetilde{\gamma},
\nonumber
\\
Z(\omega)\sqrt{\omega^2-\Delta^2(\omega)}&=&\widetilde{\Omega}+i\widetilde{\Gamma}.
\end{eqnarray*}
Here and in what follows, the branch of the square root is chosen in such a way that the sign of $\widetilde{\Omega}$ is the same as the sign of $\omega$. Furthermore we assume that, with this sign convention, $\widetilde{\Gamma}>0$. As will become clear soon, $\widetilde{\Omega}$ is the energy of the quasiparticle in the superconducting state and $\widetilde{\Gamma}$ is its lifetime. The quantities $\widetilde{\omega}$ and $\widetilde{\gamma}$ have a similar but less transparent meaning. 

Let us also introduce the following notation for a Lorentzian with width $\widetilde{\Gamma}$:
$$
\delta_{\widetilde{\Gamma}}(x)=\frac{1}{\pi}
\frac{\widetilde{\Gamma}}{x^2+\widetilde{\Gamma}^2}.
$$ 
 
In Appendix~C of \cite{Herman17a} it has been shown that the spectral function of a general Eliashberg superconductor, when viewed as a function of energy $\varepsilon$ at fixed frequency $\omega$, takes the following simple form
\begin{eqnarray}
A_s(\varepsilon,\omega)&=&
P\delta_{\widetilde{\Gamma}}(\varepsilon-\widetilde{\Omega})
+Q\delta_{\widetilde{\Gamma}}(\varepsilon+\widetilde{\Omega})
\nonumber
\\
&+&
R\frac{4\pi\widetilde{\Omega}^2}{\widetilde{\Gamma}}
\delta_{\widetilde{\Gamma}}(\varepsilon-\widetilde{\Omega})
\delta_{\widetilde{\Gamma}}(\varepsilon+\widetilde{\Omega}),
\label{eq:momentum}
\end{eqnarray}
where the $\omega$-dependent weights of the three terms are
$$
P=\frac{1}{2}\left(\frac{\widetilde{\gamma}}{\widetilde{\Gamma}}+1\right)\!,
\,\,
Q=\frac{1}{2}\left(\frac{\widetilde{\gamma}}{\widetilde{\Gamma}}-1\right)\!,
\,\,
R=\frac{1}{2}
\left(\frac{\widetilde{\omega}}{\widetilde{\Omega}}-\frac{\widetilde{\gamma}}{\widetilde{\Gamma}}\right).
$$

In what follows, the product of the two Lorentzians which appears in Eq.~\eqref{eq:momentum} will be approximated by the function
$$
\delta_{\widetilde{\Gamma}}(\varepsilon-\widetilde{\Omega})
\delta_{\widetilde{\Gamma}}(\varepsilon+\widetilde{\Omega})
\approx 
\frac{1}{4\pi}\frac{\widetilde{\Gamma}}{{\widetilde{\Omega}}^2+{\widetilde{\Gamma}}^2}
\left[
\delta_{\widetilde{\Gamma}}(\varepsilon-\widetilde{\Omega})
+\delta_{\widetilde{\Gamma}}(\varepsilon+\widetilde{\Omega})
\right].
$$
Note that the areas under the curves of $\varepsilon$ on both sides of this approximate equality are the same for all ratios of $\widetilde{\Omega}$ and $\widetilde{\Gamma}$. As pointed out in \cite{Herman17a}, the approximation is very good in the case when $\widetilde{\Omega}\gg \widetilde{\Gamma}$. In the opposite case $\widetilde{\Omega}\ll \widetilde{\Gamma}$ the approximation is still reasonable, although not perfect.

Plugging this approximation into Eq.~\eqref{eq:momentum}, after some algebra we find the following simplified expression for the spectral function
$$
A_s(\varepsilon,\omega)=
\frac{1}{2}\left[n(\omega)+1\right]
\delta_{\widetilde{\Gamma}}(\varepsilon-\widetilde{\Omega})+
\frac{1}{2}\left[n(\omega)-1\right]
\delta_{\widetilde{\Gamma}}(\varepsilon+\widetilde{\Omega}),
$$
according to which the spectral function consists of just two Lorentzians with weights determined by the function
\begin{equation}
n(\omega)={\rm Re}\left[\frac{\widetilde{\omega}+i\widetilde{\gamma}}
{\widetilde{\Omega}+i\widetilde{\Gamma}}\right]
={\rm Re}\frac{\omega}{\sqrt{\omega^2-\Delta^2(\omega)}}.
\label{eq:dos}
\end{equation}

Now let us insert the simplified spectral function into the defining Eq.~\eqref{eq:dos_definition} for the density of states. If we take into account that the width of the Lorentzians is $\widetilde{\Gamma}$, we find immediately that
$$
N_s(\omega)=
\frac{1}{2}\left[n(\omega)+1\right] {\overline N_{0}}(\widetilde{\Omega})
+
\frac{1}{2}\left[n(\omega)-1\right] {\overline N_{0}}(-\widetilde{\Omega}),
$$
where ${\overline N_{0}}(\varepsilon)$ is an appropriately smoothened version of the auxiliary function $N_0(\varepsilon)$. 

More formally, the above expression for $N_s(\omega)$ can be obtained as follows. First, let us introduce an analytic function ${\cal N}(z)$ in the upper half-plane of complex energy $z$ whose real part reduces to the function $N_0(\varepsilon)$ at the real axis, ${\rm Re}\left[{\cal N}(\varepsilon+i0)\right]=N_0(\varepsilon)$. Next we complete the $\varepsilon$ integral in Eq.~\eqref{eq:dos_definition} by a semicircle at infinity in the upper half-plane, obtaining a closed path $C$. Since $A_s(z,\omega)\propto z^{-2}$ at the semicircle, Eq.~\eqref{eq:dos_definition} can be rewritten as
$$
N_s(\omega)={\rm Re}\int_C dz {\cal N}(z) A_s(z,\omega).
$$
Making use of the residue theorem one can check readily that our expression for $N_s(\omega)$ is valid and ${\overline N_{0}}(\varepsilon)$ can be calculated as ${\overline N_{0}}(\varepsilon)={\cal N}(\varepsilon+i\widetilde{\Gamma})$.

The same set of arguments applied to the normal state would lead to the expression $N_n(\omega)={\overline N_0}(\widetilde{\omega})$, since $n(\omega)=1$ in this case. Strictly speaking, in the normal state the width of the Lorentzians is $\widetilde{\gamma}$ and not $\widetilde{\Gamma}$ as in the superconducting state, and therefore we should introduce two different functions ${\overline N_{0}}(\varepsilon)$ for the normal and superconducting states. However, in what follows we will neglect this difference. An a posteriori check of the quality of our approximations will be presented in Appendix~\ref{sec:test}.

Finally, in the low-energy limit (with respect to the phonon energy) we can write $\widetilde{\omega}\approx (1+\lambda)\omega$ where $\lambda$ is the electron-phonon coupling constant. Therefore the auxiliary function ${\overline N_0}(\omega)$ can be expressed in terms of the observable density of states in the normal state, $N_n(\omega)$, as ${\overline N_0}(\omega)=N_n(\tfrac{\omega}{1+\lambda})$. 

In what follows it will be useful to split the density of states $N_i(\omega)$ in the normal and superconducting states, $i=n,s$, into even and odd components. To this end, we write $N_i(\omega)=N_{i e}(\omega)+N_{i o}(\omega)$, where $N_{i e}(-\omega)=N_{i e}(\omega)$ and $N_{i o}(-\omega)=-N_{i o}(\omega)$. Rewriting the expression for $N_s(\omega)$ in terms of the observable normal-state density of states $N_n(\omega)$, it can be finally written as
\begin{eqnarray}
N_{se}(\omega)&=&n(\omega) N_{ne}(\Omega),
\nonumber
\\
N_{so}(\omega)&=&N_{no}(\Omega),
\label{eq:SNdos}
\end{eqnarray}
where $\Omega=\widetilde{\Omega}/(1+\lambda)$. But once it is established that the superconducting density of states at energy $\omega$ depends on the normal-state density of states at energy $\Omega$ via Eq.~\eqref{eq:SNdos}, the conservation of the total number of states implies that $n(\omega)=d\Omega/d\omega$ must hold, since the odd part of the density of states does not contribute to the total number of states.  Therefore we can finally write
\begin{equation}
N_{se}(\omega)=\frac{d\Omega}{d\omega} N_{ne}(\Omega).
\label{eq:change}
\end{equation}

The result Eq.~\eqref{eq:change} simply means that the density of states in the superconducting state at energy $\omega$ is determined by the normal-state density of states at a single $\omega$-dependent energy $\Omega(\omega)$, to be determined later. If we introduce the total number of states $H_i(\omega)=\int_0^\omega d\nu N_i(\nu)$ in the phases $i=s,n$ with energies less than $\omega$, then Eq.~\eqref{eq:change} can be reinterpreted as the requirement $H_s(\omega)=H_n(\Omega)$. It is the obvious that the equality of two different growing functions $H_i$ has to be realized via a change of the scale $\Omega=\Omega(\omega)$.

The function $n(\omega)=d\Omega/d\omega$ will be called the dos-function in what follows, since Eq.~\eqref{eq:dos} demonstrates that it plays the role of the density of states of a hypothetical superconductor with a constant density of states in the normal state.

\subsection{Application to model data}
Before proceeding, let us first investigate how $N_s(\omega)$ described by Eqs.~(\ref{eq:SNdos},\ref{eq:change}) depends on the form of the normal-state density of states $N_n(\omega)$. In order to deal with a simple instructive model, we assume that the function $\Omega(\omega)$ is given by $\Omega(\omega)={\rm Re}\sqrt{(\omega+i\Gamma)^2-\Delta^2}$ , which implies that the dos-function acquires the Dynes form \cite{Kavicky20} $n(\omega)={\rm Re}[(\omega+i\Gamma)/\sqrt{(\omega+i\Gamma)^2-\Delta^2}]$. 

\begin{figure}[t]
  \includegraphics[width = 7.5 cm]{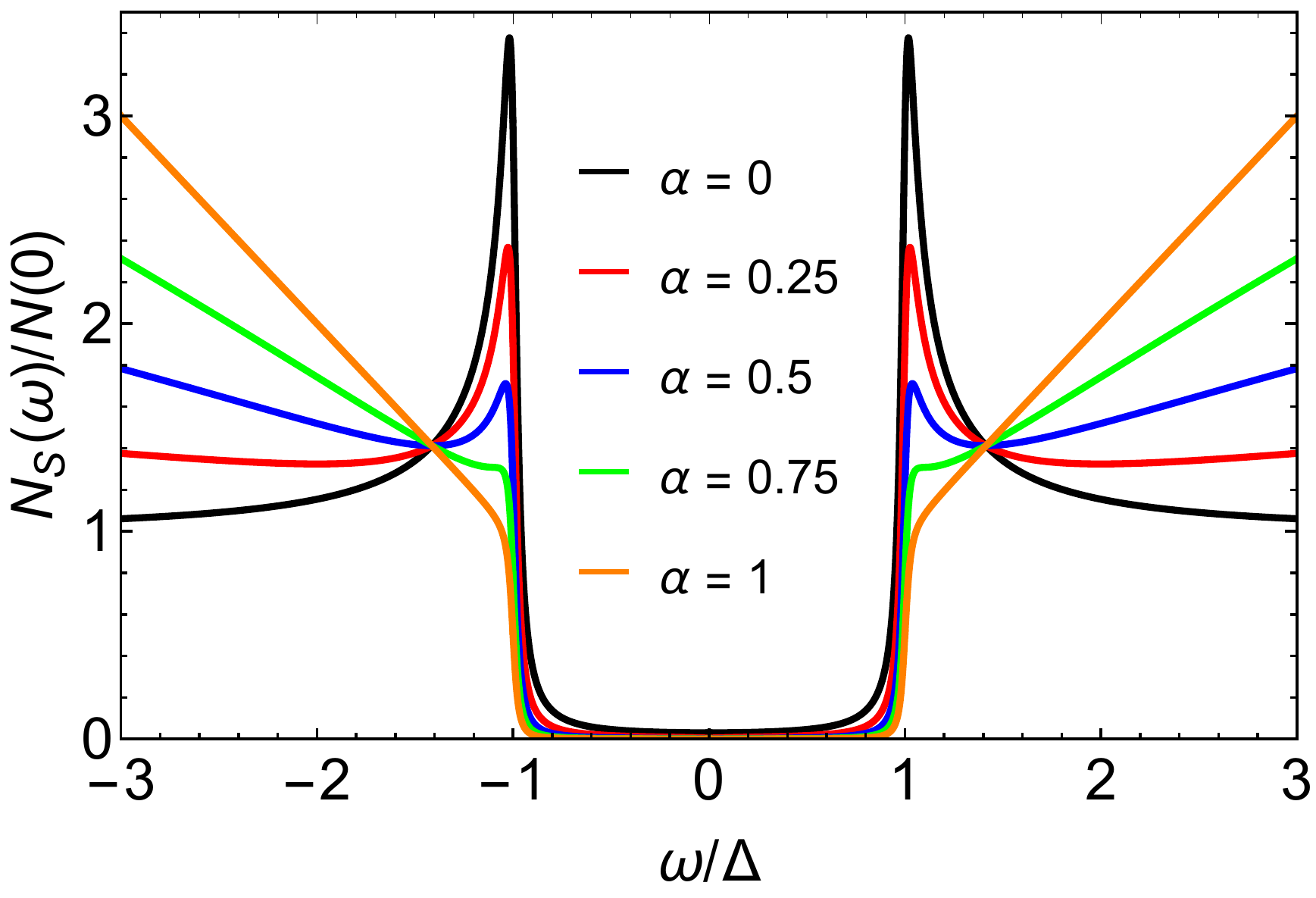}
\caption{Prediction of Eq.~\eqref{eq:change} for the superconducting density of states $N_s(\omega)$ of a Dynes superconductor with a small pair-breaking rate $\Gamma=0.03~\Delta$. The normal-state density of states is described by the particle-hole symmetric model $N_n(\omega)=(|\omega|/\Delta)^\alpha N(0)$ with several values of the exponent $\alpha$.}
\label{fig:schematic}
\end{figure}

In Fig.~\ref{fig:schematic} we start by showing the superconducting density of states $N_s(\omega)$ calculated for a simple particle-hole symmetric model of the normal state $N_n(\omega)=(|\omega|/\Delta)^\alpha N(0)$ and several values of the exponent $\alpha$. Note that while the apparent gap essentially does not depend on $\alpha$, the magnitude of the "coherence peaks" in the vicinity of $\pm\Delta$ quickly decreases with increasing $\alpha$. 

\begin{figure}[t]
  \includegraphics[width = 7.5 cm]{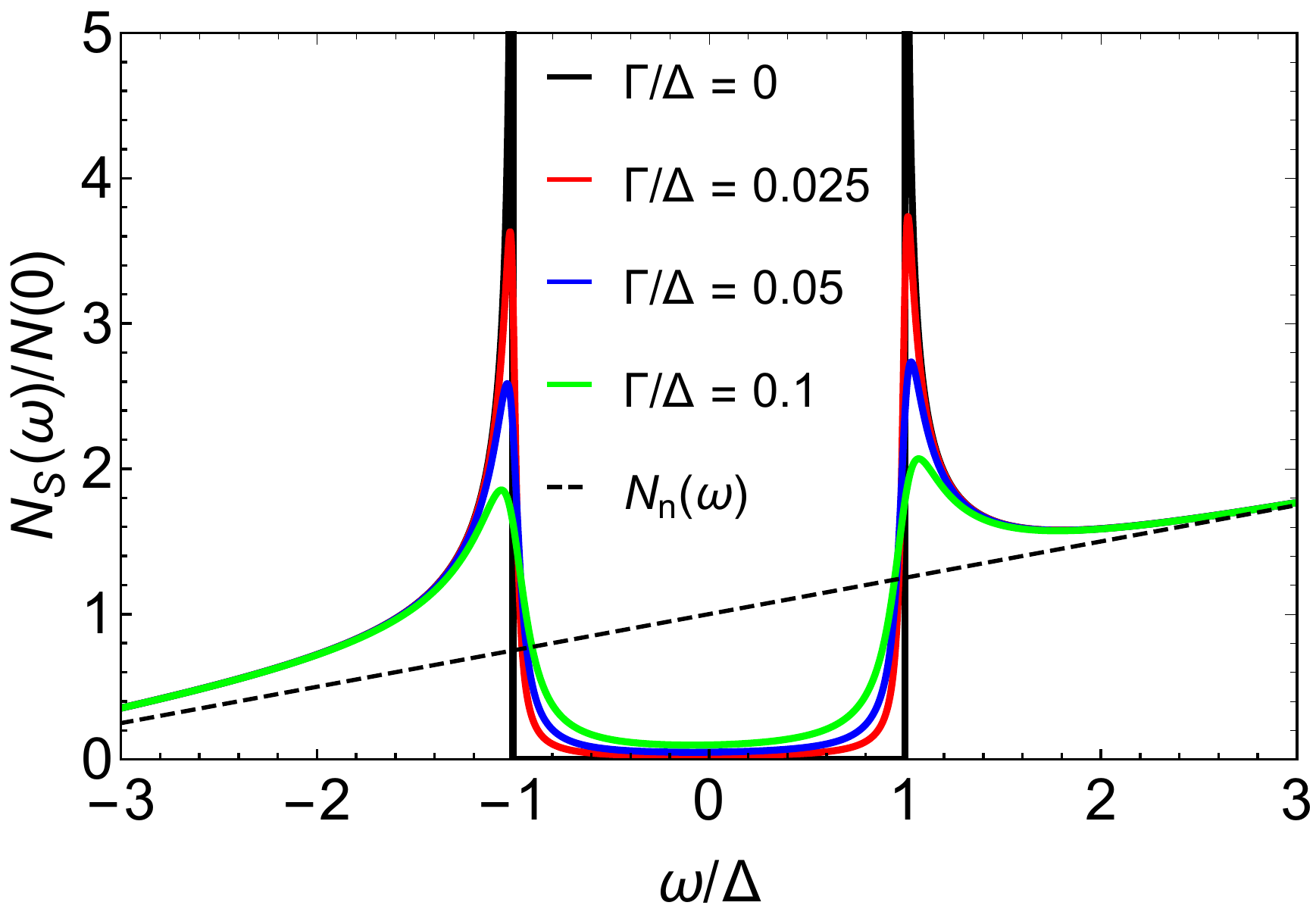}
\caption{Prediction of Eq.~\eqref{eq:change} for the superconducting density of states $N_s(\omega)$ of Dynes superconductors with several values of the pair-breaking rate $\Gamma$. The normal-state density of states $N_n(\omega)$, which is not particle-hole symmetric, is shown by the dashed line.}
\label{fig:schematic_asymmetry}
\end{figure}

On the other hand, in Fig.~\ref{fig:schematic_asymmetry} we study a model system with a normal state that is not particle-hole symmetric. In this case we find that the asymmetry of $N_s(\omega)$ is comparable to that of $N_n(\omega)$ only at energies much larger than the gap, while inside the gap it is substantially reduced, as was to be expected.

The emergent granularity scenario emphasizes that both, the normal and the superconducting densities of states, $N_n(\omega)$ and $N_s(\omega)$, are position-dependent. If we interpret Eqs.~(\ref{eq:SNdos},\ref{eq:change}) as a relation between position-dependent quantities, then one can study also the spatial variations of the dos-function $n(\omega)$. One of the important questions which can be asked within such approach has to do with the origin of the observed spatial modulations of $N_s(\omega)$: are they predominantly caused by the spatial modulations of $n(\omega)$, as is routinely assumed \cite{Feigelman10}, or is the spatial modulation of $N_n(\omega)$ more important?

As shown in Fig.~\ref{fig:schematic}, the frequently observed spatially constant gap coexisting with spatially modulated coherence peaks (see, e.g., Fig.~3 in \cite{Postolova20}) may in fact be caused by the latter mechanism, namely by the spatial modulation of $N_n(\omega)$. Also note that the suppressed coherence peaks do not necessarily imply the presence of localized Cooper pairs as is sometimes claimed \cite{Sacepe11}. 

\subsection{Application to thin TiN films}
Let us turn back to our main goal of extracting the gap function $\Delta(\omega)$ from experimental data. Our key observation is that Eq.~\eqref{eq:change} can be regarded as a first-order differential equation for the unknown function $\Omega=\Omega(\omega)$: 
\begin{equation}
\frac{d\Omega}{d\omega}=\frac{N_{se}(\omega)}{N_{ne}(\Omega)},
\label{eq:diff_eq}
\end{equation}
with initial condition $\Omega(0)=0$. Of course, in order to determine the right-hand side of Eq.~\eqref{eq:diff_eq},  both the normal and the superconducting densities of states will have to be known.  Once the function $\Omega(\omega)$ is found, the dos-function can be determined easily from $n(\omega)=d\Omega/d\omega$.  

In order to demonstrate the feasibility of this procedure, in what follows it shall be applied to the data for a homogeneously disordered thin film of TiN reported in Fig.~12 of \cite{Postolova20}. Since the experiment is performed at low temperatures, we can neglect the difference between the measured tunneling conductance and the density of states. This assumption is supported also by our preliminary attempt to explicitly eliminate the effect of the finite value of the temperature which leads to only very minor changes of our results. 

Our crucial assumption, supported by Fig.~2 of \cite{Postolova20}, is that the non-trivial form of the normal-state density of states $N_n(\omega)$ is not caused by superconducting fluctuations. In fact, the normal-state corrections reported in \cite{Postolova20} are slightly larger in the field of 7~T than in 4~T, while both fields are much larger than the estimated critical field of 2.65~T. Therefore it is very unlikely that the corrections are due to superconducting fluctuations. The extraction of the normal-state density of states at zero field from the data presented in Fig.~12 of \cite{Postolova20} is described in detail in Appendix~\ref{sec:normal}.

The next subtle point has to do with the fact that tunneling determines the density of states only up to a multiplicative constant, and therefore the relative normalization of the functions $N_{se}(\omega)$ and $N_{ne}(\omega)$ is unknown in general. These two functions should merge in the limit of energies which are much larger than the superconducting gap, but since  experimental data is available from \cite{Postolova20} only up to $|\omega|=\Lambda=1.1$~meV, a different procedure has to be used instead. 

In this work, the relative norm of $N_{se}(\omega)$ and $N_{ne}(\omega)$ is chosen so that the dos-function $n(\omega)=d\Omega/d\omega$ which is implied by the solution of Eq.~\eqref{eq:diff_eq} satisfies the constraint
\begin{equation}
\int_{-\infty}^\infty d\omega [n(\omega)-1]=0,
\label{eq:constraint}
\end{equation}
which requires that the superconducting transition conserves the total number of states. Moreover, this constraint guarantees that, as $\omega$ increases, the difference between $\Omega(\omega)$ and $\omega$ vanishes. 

In order to estimate the contribution to the integral in Eq.~\eqref{eq:constraint} of the region $|\omega|>\Lambda$ where no data is available, in this region we assume that $n(\omega)=1+a/\omega^2+b/\omega^4$. The two fitting parameters $a$ and $b$ are determined by requiring (i) that $n(\omega)$ is continuous at $\omega=\Lambda$, and (ii) that the constraint Eq.~\eqref{eq:constraint} is satisfied. It turns out that $a$ and $b$ can be found in a finite range of relative normalizations of the functions $N_{se}(\omega)$ and $N_{ne}(\omega)$, see Appendix~\ref{sec:normalization}. The optimal normalization is then chosen by requiring that also the derivative of $n(\omega)$ is continuous at $\omega=\Lambda$. 

\begin{figure}[t]
  \includegraphics[width = 8.6 cm]{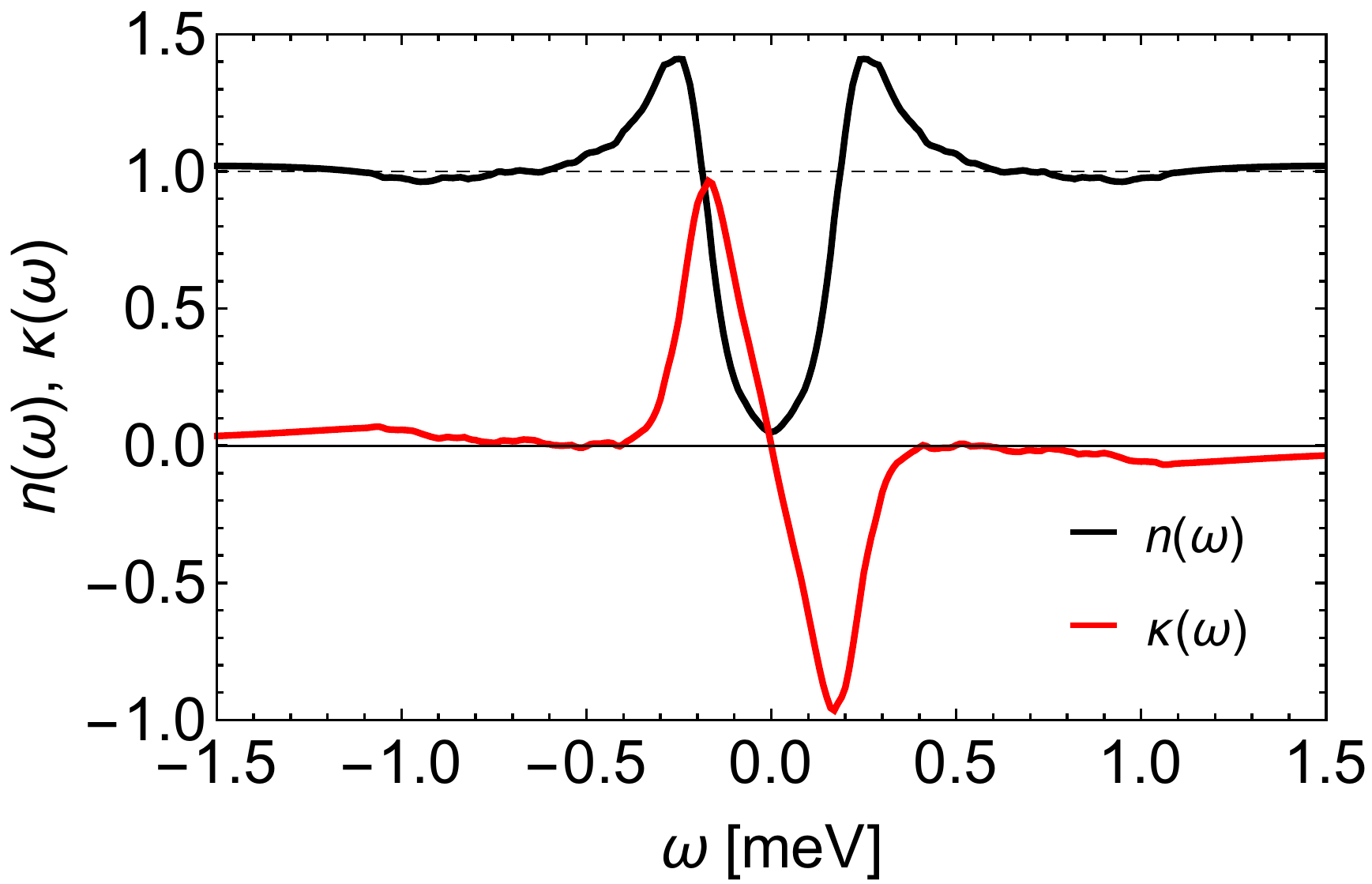}
\caption{The dos-function $n(\omega)$ determined by solving the symmetric part of Eq.~\eqref{eq:change}. $N_{se}(\omega)$ is determined by symmetrization of the experimental data from \cite{Postolova20} and $N_{ne}(\omega)$ is taken from Fig.~\ref{fig:normal_extrapol} in Appendix~\ref{sec:normal}. The relative normalization of $N_{se}(\omega)$ and $N_{ne}(\omega)$ has been determined as described in the text. The high-energy prolongation by $n(\omega)=1+a/\omega^2+b/\omega^4$ with $a\approx 0.104$~meV$^2$ and $b\approx-0.132$~meV$^4$ is plotted as well. Also shown is $\kappa(\omega)$, the Kramers-Kronig partner of $n(\omega)$.}
\label{fig:dOmegadomega}
\end{figure}

The resulting dos-function $n(\omega)$ is plotted in Fig.~\ref{fig:dOmegadomega}. It looks much more BCS-like than the superconducting density of states $N_s(\omega)$. In particular, the coherence peaks are substantially amplified. Note that this should have been expected already based on the results of the model calculation presented in Fig.~\ref{fig:schematic}. 

However, the extracted dos-function $n(\omega)$ is not completely BCS-like. The largest qualitative difference with respect to the BCS prediction is related to the presence of broad local minima of $n(\omega)$ at $|\omega|=\omega_\ast\approx 1$~meV. We find that this is a robust feature which is present in a finite range of relative normalizations of $N_{se}(\omega)$ and $N_{ne}(\omega)$, see Appendix~\ref{sec:normalization}.

In what follows, it will be useful to view the dos-function $n(\omega)$ as the real part of a complex function $n(\omega)+i\kappa(\omega)$ where $\kappa(\omega)$ is the Kramers-Kronig partner of $n(\omega)$, since the continuation of the function $n(\omega)+i\kappa(\omega)$ to the upper half-plane of complex frequencies is an analytic function. The imaginary part $\kappa(\omega)$ can be easily found by integration and the result is shown in Fig.~\ref{fig:dOmegadomega}. Note that close to the same energies $|\omega|\approx \omega_\ast$ where the local minima of $n(\omega)$ are situated, there exist also peaks of $|\kappa(\omega)|$, as required by the Kramers-Kronig relations. Again, these features are present in a finite range of relative normalizations of $N_{se}(\omega)$ and $N_{ne}(\omega)$, see Appendix~\ref{sec:normalization}.

\section{Extraction of the gap function} 
\label{sec:delta}
Our next goal is to invert Eq.~\eqref{eq:dos} and to determine the complex gap function $\Delta(\omega)=\Delta_1(\omega)+i\Delta_2(\omega)$ from the already known dos-function $n(\omega)$. We assume that $\Delta(-\omega)=\Delta^\ast(\omega)$ as in the standard Eliashberg theory, although we emphasize that the self-consistent Born approximation on which the Eliashberg theory is based might not be sufficient to describe the nearly localized superconductors \cite{note_born}. A similar conclusion regarding the validity of the Eliashberg theory has been reached long ago also on purely experimental grounds \cite{Dynes86}.

\subsection{Model gap function} 
Before proceeding with the actual extraction of $\Delta(\omega)$ from $n(\omega)$, we would like to point out that both of the main qualitative deviations of the observed complex dos-function from the BCS prediction, i.e. the local minima of $n(\omega)$ at $|\omega|\approx \omega_\ast$ and the peaks of $|\kappa(\omega)|$ close to the same energy, can be explained by a simple model gap function $\Delta(\omega)$. 

In fact, let us assume that the real part of the gap function is similar to a simple two-value function, with a smaller gap $\Delta_1(\omega)\approx \Delta_0$ at small energies $|\omega|<\omega_\ast$ and a larger gap $\Delta_1(\omega)\approx \Delta_\infty$ for $|\omega|>\omega_\ast$. Making use of the results of \cite{Bevilacqua16}, a smooth complex function with these properties can be found, which satisfies also the Kramers-Kronig relations. The result reads as
\begin{eqnarray}
\Delta(\omega)&=&\Delta_\infty+\left(\Delta_0 - \Delta_\infty\right)F(\omega),
\label{eq:model_delta}
\\
F(\omega)&=&\frac{i}{\pi}\left[\Psi\left(\frac{1}{2}+\frac{\omega+\omega_{*}}{2\pi i\Theta}\right)-\Psi\left(\frac{1}{2}+\frac{\omega-\omega_{*}}{2\pi i\Theta}\right)\right],
\nonumber
\end{eqnarray}
where $\Psi(z)$ is the digamma function and $\Theta$ measures the width of the transition regions around $|\omega|=\omega_\ast$. 

\begin{figure}[t]
  \includegraphics[width = 7.5 cm]{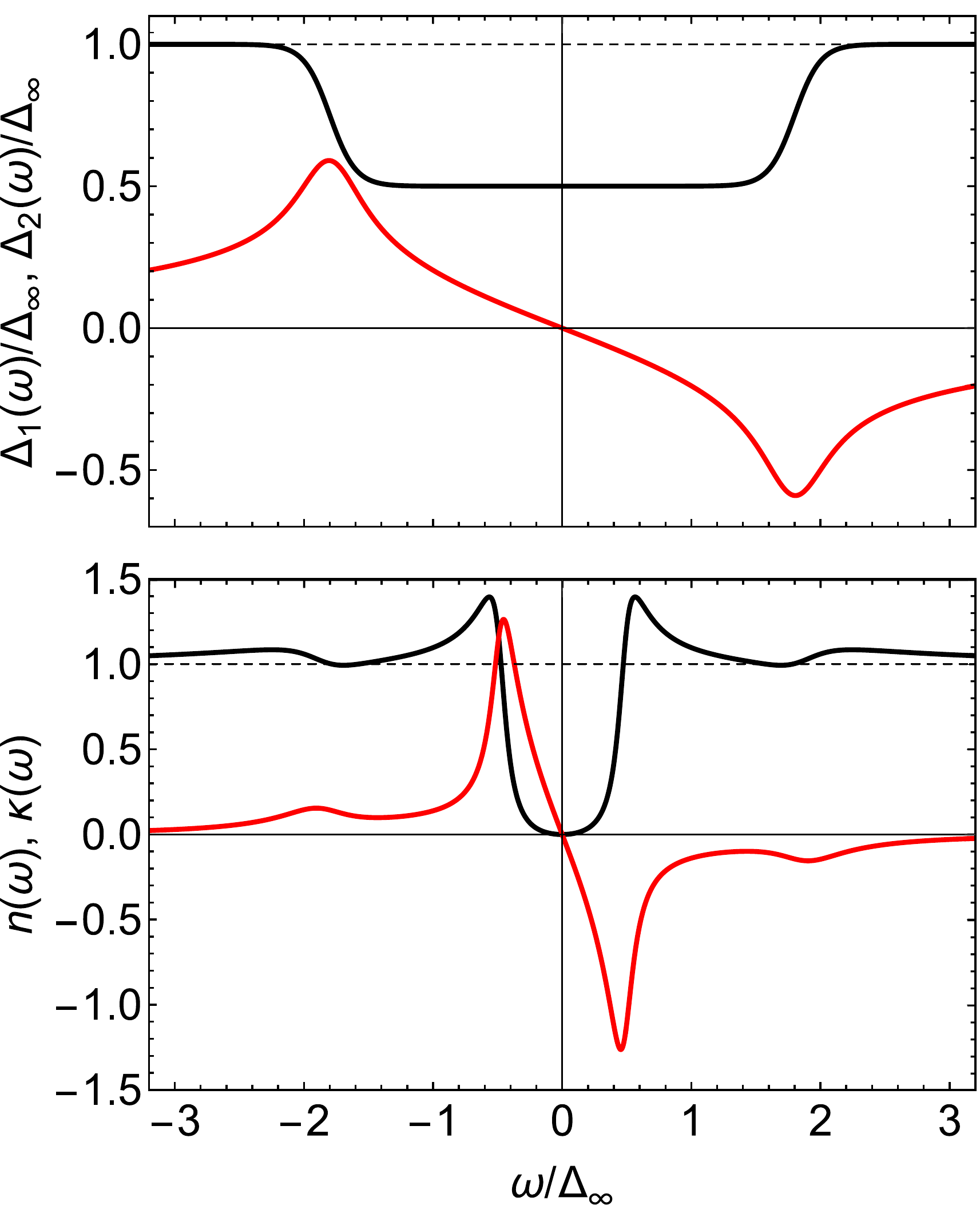}
\caption{Upper panel: the model gap function $\Delta(\omega)$ described by Eq.~\eqref{eq:model_delta} for $\Delta_0/\Delta_\infty=0.5$, $\omega_\ast/\Delta_\infty=1.8$, and $\Theta/\Delta_\infty=0.1$. Lower panel: the complex dos-function implied by this $\Delta(\omega)$. In both panels, the black (red) lines correspond to the real (imaginary) parts of the functions.}
\label{fig:model_delta}
\end{figure}

As shown in Fig.~\ref{fig:model_delta}, this model gap function does produce the qualitative features of the measured dos-function $n(\omega)+i\kappa(\omega)$. Thus we should expect that the numerical solution for $\Delta(\omega)$ exhibits at least some similarity to the model Eq.~\eqref{eq:model_delta}.

\subsection{Direct extraction}
Let us proceed by extracting the gap function $\Delta(\omega)$ directly from the measured data. To this end, we will make use of a procedure proposed in \cite{Galkin74}, slightly modified for the case of gapless superconductors. The crucial observation is that the complex gap function $\Delta(\omega)$ can be obtained from the algebraic expression
\begin{equation}
\Delta^2(\omega)=\omega^2\left[1-1/(n+i\kappa)^2\right].
\label{eq:gap_function}
\end{equation}
In fact, when continued from $\omega$ to the upper half-plane of the complex frequencies $z$, the right-hand side of Eq.~\eqref{eq:gap_function} can be easily seen to be analytic, since $n+i\kappa$ is analytic and the inequality $n(z)>0$ holds in the whole upper half-plane. Therefore Eq.~\eqref{eq:gap_function}  is compatible with the requirement that $\Delta(z)$ is analytic in the upper half-plane.

\begin{figure}[t]
\includegraphics[width = 7.5 cm]{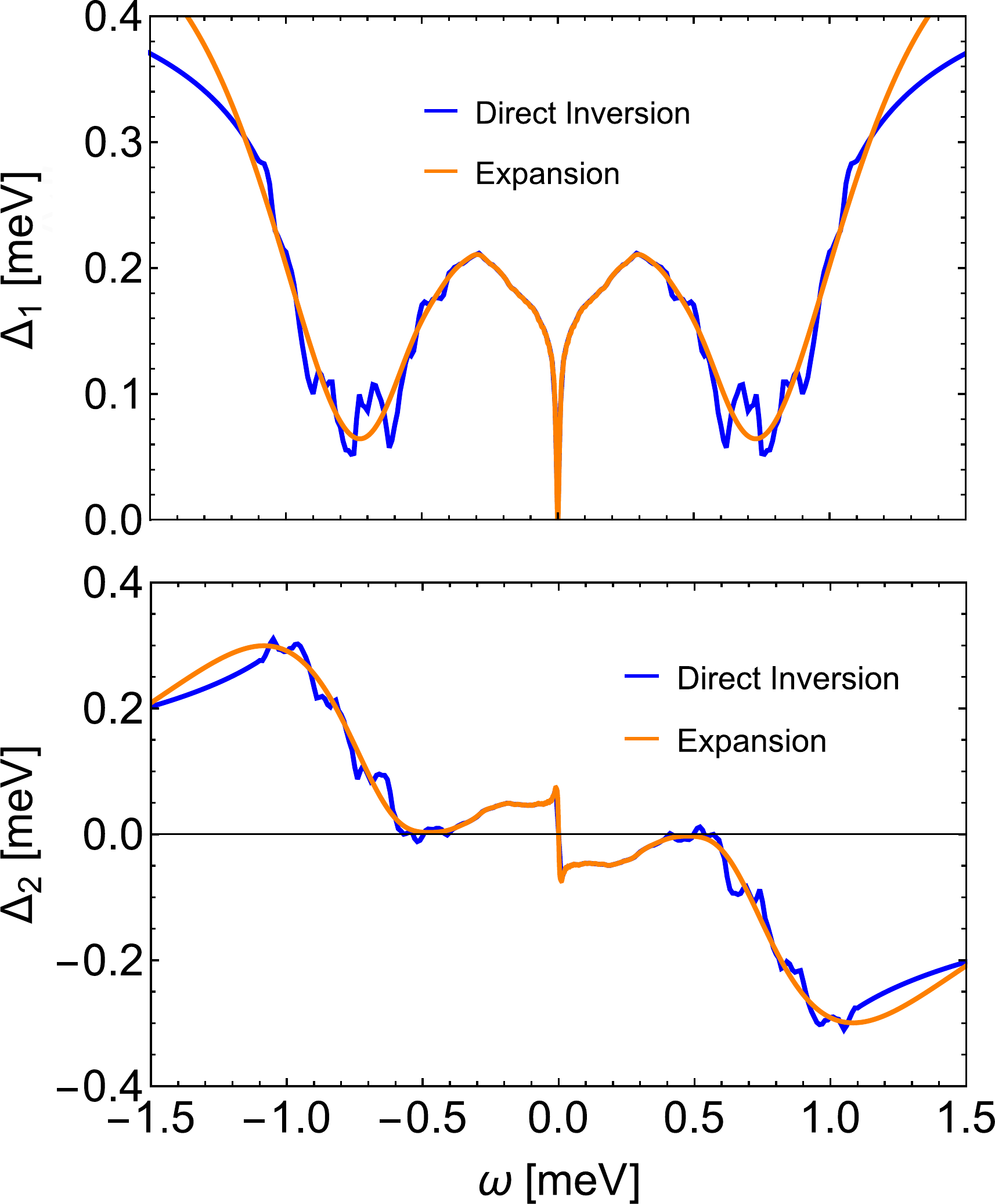}
\caption{Real and imaginary parts of the gap function extracted from the dos-function plotted in Fig.~\ref{fig:dOmegadomega}. Blue (orange) line: result of the direct inversion (inversion by expansion with $N=35$ and $E=0.1$~meV).}
\label{fig:delta}
\end{figure}

The result of the direct inversion via Eq.~\eqref{eq:gap_function}, where the complex dos-function is taken from Fig.~\ref{fig:dOmegadomega}, is shown in Fig.~\ref{fig:delta}. 

The potentially troublesome point of the presented  analysis is that, at a fixed frequency $\omega$, the complex number $\Delta(\omega)$ is determined only up to an overall sign. Therefore the choice of signs at two different frequencies is completely independent of each other.

Fortunately, in regions where the absolute value $|\Delta(\omega)|$ is large, this does not cause any problems. In fact, having fixed the signs of the real and imaginary parts $\Delta_1$ and $\Delta_2$ at some frequency, due to the expected continuity of $\Delta(\omega)$ we have to take the same sign convention also at the neighboring frequency. 

Problems therefore arise only in regions where $|\Delta(\omega)|$ is small, because in these regions continuity does not tell us how to match the sign conventions at neighboring frequencies. Thus, if such regions are present, then we are not guaranteed that the function $\Delta(\omega)$ is reconstructed correctly. 

There does exist a tool, however, which makes it possible to decide whether the signs have been chosen correctly. Namely, the complex function $\Delta(\omega)$ should obey the Kramers-Kronig relations. We have checked that our choice of signs presented in Fig.~\ref{fig:delta} does satisfy these relations. 

It is worth pointing out in passing that, when calculating $\Delta_1(\omega)$ from $\Delta_2(\omega)$ using the Kramers-Kronig relations, we had to add by hand the asymptotic value of $\Delta_\infty=\sqrt{2a}$ implied by the functional form of the prolongation of $n(\omega)$ to large frequencies. For completeness let us also add that at high energies we find $\Delta_2(\omega)\propto \omega^{-1}$ which is consistent with $\kappa(\omega) \propto \omega^{-3}$.

Yet another remark is in place at this point. Although formally our solution for the gap function stays finite as $\omega$ grows to large values, this is only an artifact. Since experimental data are available only up to $\Lambda$, we can not make any conclusions on the behavior of $\Delta(\omega)$ at energies much larger than $\Lambda$. 

\subsection{Extraction by expansion} 
As is obvious from the comparison of Fig.~\ref{fig:dOmegadomega} with Fig.~\ref{fig:delta}, the direct extraction of $\Delta(\omega)$ using Eq.~\eqref{eq:gap_function} strongly amplifies the noise which is inevitably present in the dos-function.  Unfortunately, one can not simply smoothen the functions $\Delta_1(\omega)$ and $\Delta_2(\omega)$, since the result would not be guaranteed to satisfy the Kramers-Kronig relations. Therefore an alternative method for the extraction of $\Delta(\omega)$ would be useful which can partly eliminate the relatively large noise of the direct extraction.

To this end, we make use of an expansion of the gap function in terms of the rational functions \cite{Weideman95}
$$
\rho_n(x)=\frac{(1+ix)^n}{(1-ix)^{n+1}}, \quad n=0,\pm 1,\pm2, \ldots,
$$
which form a complete basis in the space of complex square-integrable functions on the real axis. The functions $\rho_n(x)$ satisfy the orthogonality relations
$$
\int_{-\infty}^\infty dx \rho_n^\ast(x)\rho_m(x)=\pi \delta_{nm},
$$
which imply that the coefficients in the expansion of the function $f(x)=\sum_n a_n\rho_n(x)$ can be calculated as 
$$
a_n=\frac{1}{\pi}\int_{-\infty}^\infty dx \rho_n^\ast(x) f(x).
$$
The first few functions $\rho_n(x)$ are shown in Appendix~\ref{sec:gap}.

The crucial point to observe is that $\rho_n(x)$ are eigenfunctions of the Hilbert transform, i.e. they satisfy
$$
\frac{1}{\pi}P\int_{-\infty}^\infty\frac{dx \rho_n(x)}{x-y}=i{\rm sgn}(n) \rho_n(y),
$$
where $P\int dx$ denotes the principal value integration \cite{Weideman95}.

Making use of the relations $\rho_{-n-1}(x)=\rho_n^\ast(x)=\rho_n(-x)$ and of the symmetry $\Delta(-\omega)=\Delta^\ast(\omega)$ we then find that the gap function can be written as
\begin{equation}
\Delta(\omega)=\Delta_\infty+2\sum_{n=0}^\infty a_n \rho_n\left(\frac{\omega}{E}\right),
\label{eq:gap_expansion}
\end{equation}
where $E$ is an arbitrary energy scale and $a_n$ are real coefficients. One can check readily that the gap function given by Eq.~\eqref{eq:gap_expansion} is analytic in the upper half-plane and satisfies the Kramers-Kronig relations. 

The coefficients $a_n$ for $n=0,1,\ldots,N$ which appear in Eq.~\eqref{eq:gap_expansion} can be determined by minimization of the cost function which is given by the distance, at energies $|\omega|<\Lambda$, between the measured complex dos-function and $n(\omega)+i\kappa(\omega)$ calculated from Eq.~\eqref{eq:gap_expansion} using the inverse of Eq.~\eqref{eq:gap_function}. The value of $\Delta_\infty$ is also taken as a variational parameter. 

The energy scale $E$ which appears in Eq.~\eqref{eq:gap_expansion} can be chosen arbitrarily. With the aim to keep the order of the expansion $N$ small, we take $E=0.1$~meV. The choice of the optimal value of $N$ for this value of $E$ is described in Appendix~\ref{sec:gap}.

In Fig.~\ref{fig:delta} we plot $\Delta(\omega)$ calculated making use of the expansion Eq.~\eqref{eq:gap_expansion}, and compare it with the result of the direct inversion. The mutual agreement between these two very different methods is seen to be very good, giving us further confidence in the results.

\subsection{Discussion}
As demonstrated in Appendix~\ref{sec:gap}, the gap functions which describe the studied TiN sample are essentially identical in the low energy region $\omega\lesssim 0.3$~meV for all acceptable relative normalizations of $N_n(\omega)$ and $N_s(\omega)$. Therefore our analysis in this energy range should be quantitatively accurate.

It is well known that in the extreme low-energy limit $\omega\rightarrow 0$, the gap function should be generically described by the Dynes formula
$$
\Delta(\omega)\approx \frac{\Delta_0 \omega}{\omega+i\Gamma}.
$$
This formula takes into account the presence of the pair-breaking processes and describes them by a frequency-independent scattering rate $\Gamma$. Both elastic \cite{Kavicky20}, and, at finite temperatures, also inelastic \cite{Mikhailovsky91} processes were shown to contribute to the finite value of $\Gamma$. 

A strong argument supporting our results presented in Fig.~\ref{fig:delta} is provided by the observation that, in agreement with general expectations, for $\omega<0.02$~meV the Dynes formula with $\Delta_0\approx 0.151$~meV and $\Gamma\approx 0.008$~meV does in fact describe our gap function $\Delta(\omega)$.

At higher energies, two robust qualitative features of $\Delta(\omega)$ are present for all acceptable relative norms of $N_{se}(\omega)$ and $N_{ne}(\omega)$, see Fig.~\ref{fig:delta} and Appendix~\ref{sec:gap}: (i) a steep increase of $\Delta_1(\omega)$ around $\omega_\ast\approx 1$~meV and (ii) a concomitant sharp negative peak of $\Delta_2(\omega)$ at the same energy. As shown by the simple model calculation, these two features are a direct consequence of the local minimum of $n(\omega)$ close to $\omega_\ast$,  which in turn is caused by the unusual suppression of $N_s(\omega)$ with respect to $N_n(\omega)$ at energies above the gap, see Appendix~\ref{sec:normalization}.

As already pointed out, the Eliashberg theory is most likely not applicable to the strongly disordered TiN thin films. In what follows we will therefore present a tentative comparison of our results with an extended version of the Eliashberg theory which takes into account not only the attractive, but also the repulsive boson-mediated interactions \cite{Bzdusek15,Sun21}. The latter type of pair-breaking forces can be due to current-current interactions, as suggested in \cite{Bzdusek15}, but also due to exchange of magnetic excitations \cite{Sun21}, etc. In principle, whether a given interaction is repulsive or attractive is governed by the type of the Pauli matrix which enters the electron-boson scattering vertex in the Nambu-Gorkov formalism.

When interpreted within such extended Eliashberg theory, the features (i) and (ii) of the extracted gap function $\Delta(\omega)$ close to $\omega_\ast$ can be seen to imply a finite coupling of the electrons to a pair-breaking mode with energy $\omega_\ast$, see Fig.~1b in \cite{Bzdusek15}. However, even if this interpretation is correct, the physical nature of the pair-breaking mode in TiN is currently unclear. The mode could be generated by the dynamical screening of the electron-electron interaction if the fermionic scenario applies, but it could also be a phase fluctuation mode of the bosonic scenario. 

\section{Conclusions}
\label{sec:conclusions}
In this paper we have introduced a methodology which allows one to extract the gap function $\Delta(\omega)$ of an s-wave superconductor, even if the normal-state density of states exhibits complicated structure due to quantum corrections and the standard inversion procedure \cite{McMillan65} therefore does not apply. As an input, our procedure requires that the tunneling conductance is known in both, the normal and the superconducting states, and in a sufficiently broad range of energies. 

The proposed procedure consists of two steps: first we extract, by means of a numerical integration of Eq.~\eqref{eq:diff_eq}, the dos-function $n(\omega)$. In the second step, the gap function $\Delta(\omega)$ is calculated from $n(\omega)$. This latter step can be dealt with by a suitable modification of the approach of \cite{Galkin74}, but also by a very efficient expansion method making use of the complete set of rational eigenfunctions of the Hilbert transform, introduced some time ago in the mathematical literature \cite{Weideman95}.

Unfortunately, in experimental papers on strongly disordered superconductors, the authors only very rarely present both, the normal and the superconducting densities of states. For instance in the influential early paper on planar tunneling into strongly disordered thin Pb films \cite{Dynes86}, only the ratio of the (thermally smeared) superconducting and normal densities of states has been presented. Many later planar tunneling studies have shown essentially the same kind of information. 

Similarly, scanning tunneling data are often reported either only in the superconducting state, or in a narrow energy range, or both. In order to demonstrate that our method does work, we have chosen to apply it to the published data for dirty TiN thin films \cite{Postolova20}, since they provide, in our opinion, the best compromise between both of the mentioned requirements. Moreover, tunneling data are available in \cite{Postolova20} also at two different magnetic fields above $H_{c2}$, which enables us to estimate the (hypothetical) normal-state density of states in the limit of low temperatures and zero field.

Applying our methodology to \cite{Postolova20}, we have demonstrated that both the dos-function $n(\omega)$ and the gap function $\Delta(\omega)$ can be succesfully extracted from the experimental data. As expected, we find that the gap function exhibits the generic Dynes form in the limit of small energies. Moreover, our results can be interpreted in terms of the coupling of the electrons to a very soft pair-breaking mode at approximately $\omega_\ast\sim$1~meV. 

The circumstance that the characteristic energy scale $\omega_\ast$ which we find is very close to the upper cut-off $\Lambda$ up to which the experimental data is available \cite{Postolova20} is definitely unfortunate. We are convinced, however, that further measurements in a broader energy range and in more finely sampled magnetic fields, both achievable with current instrumentation, can substantially eliminate the uncertainties of the present analysis.

To conclude, we believe that our method for extracting the gap function can provide crucial information on possible changes of the pairing glue and/or of the pair-breaking processes as the QBS critical point is approached. Moreover, spatial variations of the gap function can be studied by the very same procedure as well. For these reasons, we believe that our methodology can be used to clarify the role played by the three main scenaria for QBS.

\section*{Acknowledgement}
This work was supported by the Slovak Research and Development Agency under Contract No.~APVV-19-0371, by the VEGA Agency under Contract No.~1/0640/20, and by the Comenius University under Contract No.~UK/343/2021. F. H. is grateful for funding from the EU Horizont 2020 under the Marie Skłodowska-Curie mechanism No. 945478.

\appendix

\section{Extraction of $N_n(\omega)$}
\label{sec:normal}
According to Fig.~12 of \cite{Postolova20}, the normal-state density of states of TiN thin films is magnetic-field dependent. However, what we need to insert into Eq.~\eqref{eq:diff_eq} is the (hypothetical) normal-state density of states at zero field. In other words, what we need is the extrapolation of the normal-state data back to $B=0$~T. 

In order to solve this problem, we are led by the observation made in \cite{Zemlicka20} that the normal-state data for MoC films in different fields merge at an energy of the order of $2\mu_B B$. Assuming that the same physics applies also to the TiN films studied in \cite{Postolova20},
we proceed as follows.

\begin{figure}[t]
  \includegraphics[width = 7.5 cm]{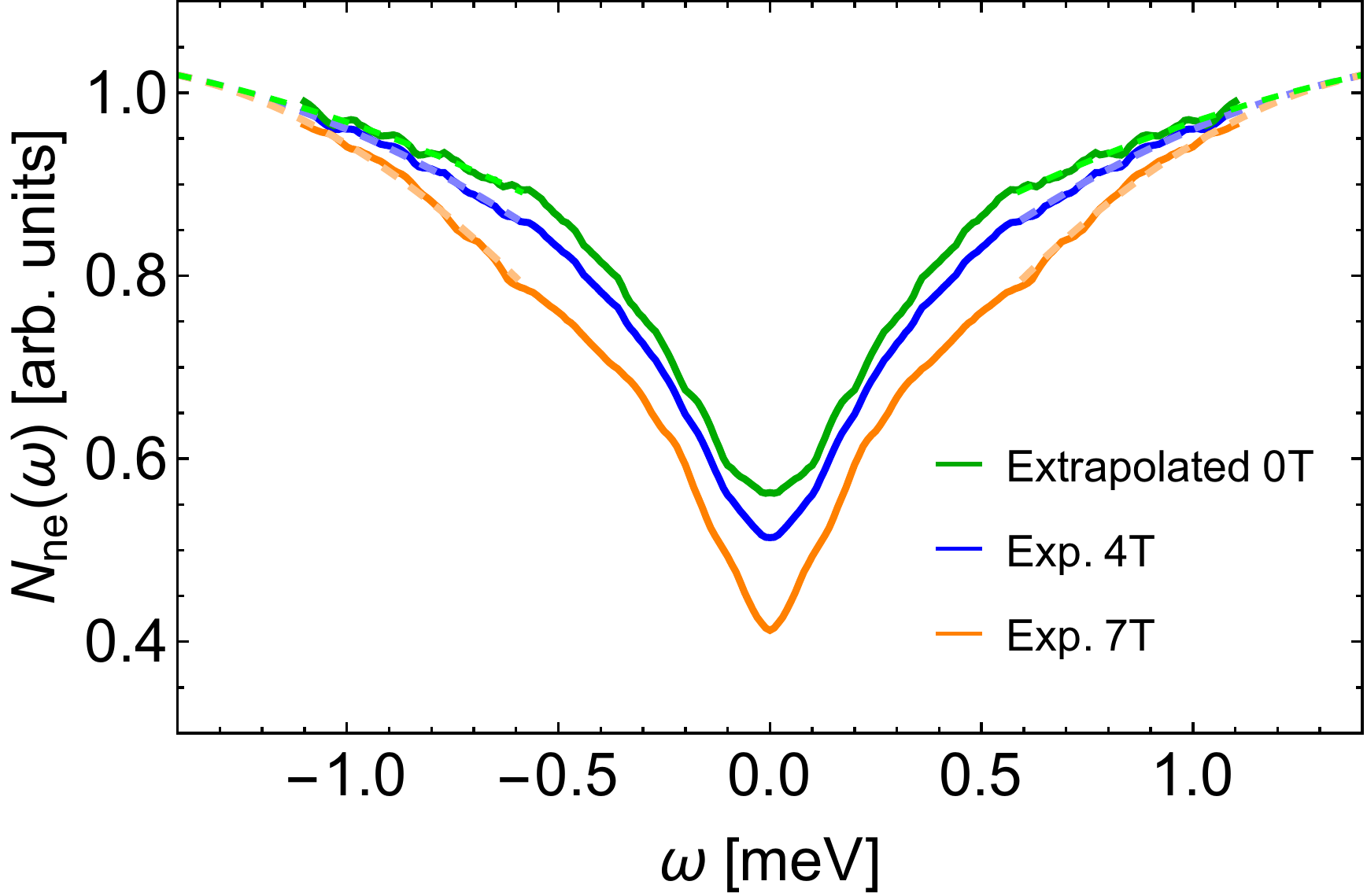}
\caption{Even part of the normal-state density of states (in arbitrary units) extrapolated from the data for 4~T and 7~T in Fig.~12 of \cite{Postolova20} to $B=0$~T. Dashed lines show smooth prolongations of the measured data which merge at $\approx$~1.4~meV.}
\label{fig:normal_extrapol}
\end{figure}

In the first step,  we choose the relative normalization of the normal-state data in 4~T and 7~T in such a way that their smooth prolongations merge at an energy outside the measured window, see Fig.~\ref{fig:normal_extrapol}. In the second step, assuming furthermore that $N_n(\omega,B)=N_n(\omega,0)+a(\omega) B^2$, we obtain an estimate of the $B=0$ density of states $N_n(\omega,0)$ from the  appropriately normalized 4~T and 7~T data, which were obtained in the first step. The result of this extrapolation procedure for the even part of $N_n(\omega,0)$ is shown in Fig.~\ref{fig:normal_extrapol}.

\begin{figure}[b]
  \includegraphics[width = 7.5 cm]{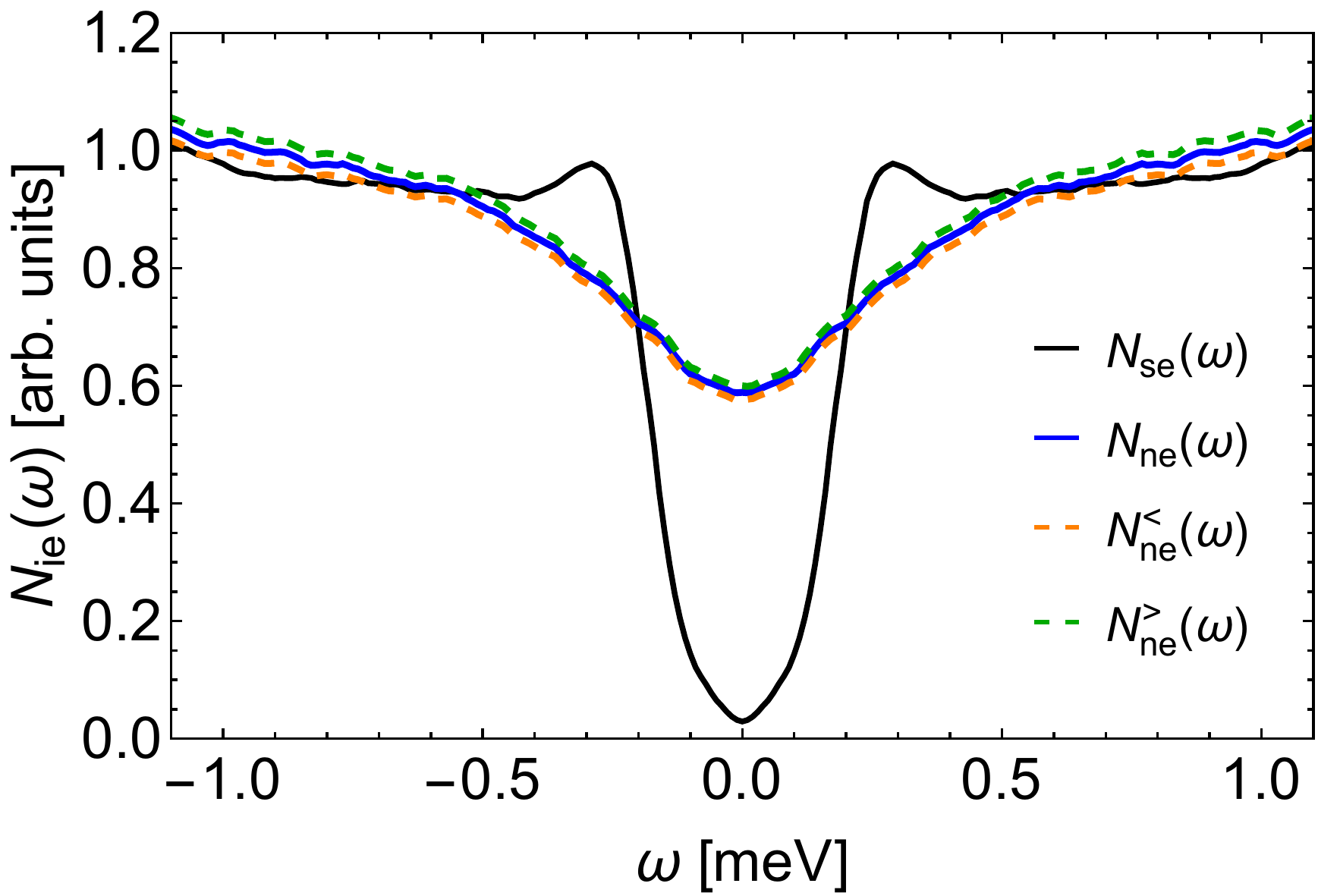}
\caption{The optimal choice $N_{ne}(\omega)$ of the normal-state density of states together with two extremal acceptable choices $N^<_{ne}(\omega)$ and $N^>_{ne}(\omega)$, see text. The norm of the superconducting density of states $N_{se}(\omega)$ is kept fixed. }
\label{fig:normalization}
\end{figure}

\section{Relative norm of $N_{se}(\omega)$ and $N_{ne}(\omega)$.}
\label{sec:normalization}
As mentioned in the main text, there is some freedom in choosing the normalization of the normal-state density of states $N_{ne}(\omega)$, if the norm of the superconducting density of states $N_{se}(\omega)$ is kept fixed. In Fig.~\ref{fig:normalization} we show three normalizations of $N_{ne}(\omega)$, all of which generate dos-functions $n(\omega)$ which satisfy the constraint Eq.~\eqref{eq:constraint} and are continuous at $|\omega|=\Lambda$. The optimal choice called $N_{ne}(\omega)$ in Fig.~\ref{fig:normalization} (and considered in the main text) leads to a solution where also the derivative $dn(\omega)/d\omega$ is smooth at $|\omega|=\Lambda$, whereas the choices $N^<_{ne}(\omega)$ and $N^>_{ne}(\omega)$ correspond to extremal solutions where the discontinuity of $dn(\omega)/d\omega$ at $|\omega|=\Lambda$ is still acceptable. 

\begin{figure}[t]
  \includegraphics[width = 7.5 cm]{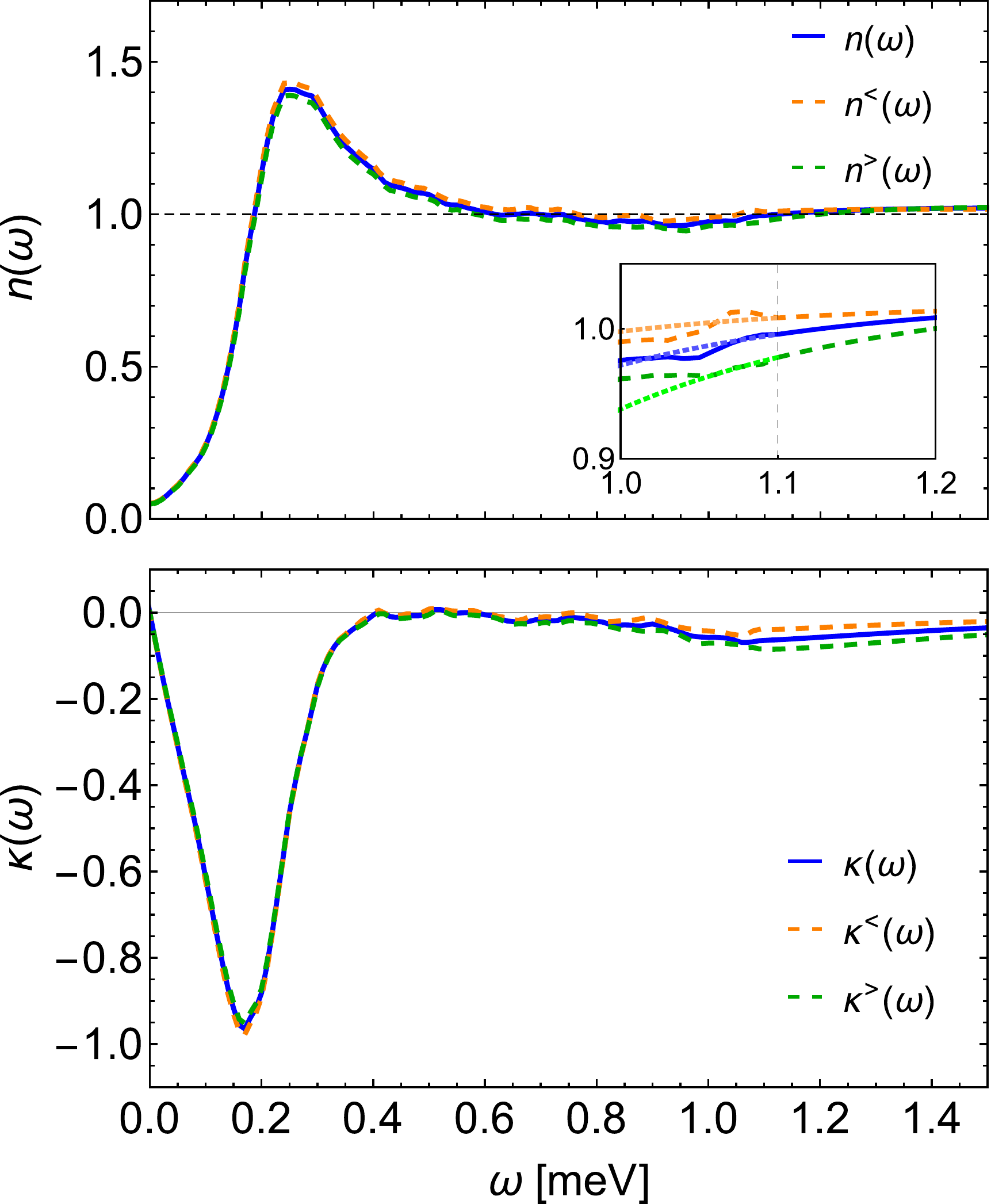}
\caption{Upper panel: The optimal dos-function $n(\omega)$ and two extremal acceptable dos-functions $n^<(\omega)$ and $n^>(\omega)$ corresponding to the choices $N_{ne}(\omega)$, $N^<_{ne}(\omega)$, and $N^>_{ne}(\omega)$ from Fig.~\ref{fig:normalization}. Only the $\omega>0$ part of these even functions is plotted. The inset shows the dos-functions in the vicinity of the energy $\omega=\Lambda=1.1$~meV (indicated by the vertical dashed line). For energies $\omega<\Lambda$ where data extracted from experiments are available, the prolongations $1+a/\omega^2+b/\omega^4$ are shown by dotted lines. Lower panel: the Kramers-Kronig partners $\kappa(\omega)$ of the three dos-functions $n(\omega)$ from the upper panel. Only the $\omega>0$ part of these odd functions is plotted.}
\label{fig:3_dos_functions}
\end{figure}

The three dos-functions $n(\omega)$, $n^<(\omega)$, and $n^>(\omega)$ extracted from the three normalizations $N_{ne}(\omega)$, $N^<_{ne}(\omega)$, and $N^>_{ne}(\omega)$ are shown in the upper panel of Fig.~\ref{fig:3_dos_functions}. The inset shows how the numerically determined dos-functions for $\omega<\Lambda$ merge with their prolongations to $\omega>\Lambda$. Note that the function $n(\omega)$ is essentially smooth at $\omega=\Lambda$, while $n^<(\omega)$ and $n^>(\omega)$ exhibit small upward and downward kinks, respectively. The coefficients describing the prolongations are $a\approx 0.069$~meV$^2$ and $b\approx -0.071$~meV$^4$ for $n^<(\omega)$ and $a\approx 0.142$~meV$^2$ and $b\approx -0.204$~meV$^4$ for $n^>(\omega)$.

The imaginary part of the complex dos-function $n(\omega)+i\kappa(\omega)$, shown in the lower panel of Fig.~\ref{fig:3_dos_functions}, is calculated by using the Kramers-Kronig relations from the three sets of data for $n(\omega)$ in the upper panel. Remarkably, all $\kappa(\omega)$ curves look quite BCS-like and they exhibit a weak structure in the vicinity of $\omega_\ast\approx 1$~meV, where $n(\omega)$ exhibits a minimum. Thanks to Eq.~\eqref{eq:constraint}, at large energies the functions $\kappa(\omega)$ decay as $\omega^{-3}$. We have also checked that, starting from $\kappa(\omega)$ and applying the Kramers-Kronig relations in the reverse direction, we do reproduce the dos-functions $n(\omega)-1$.

\section{Extraction of $\Delta(\omega)$ from $n(\omega)+i\kappa(\omega)$}
\label{sec:gap}
\begin{figure}[t]
\includegraphics[width = 7.5 cm]{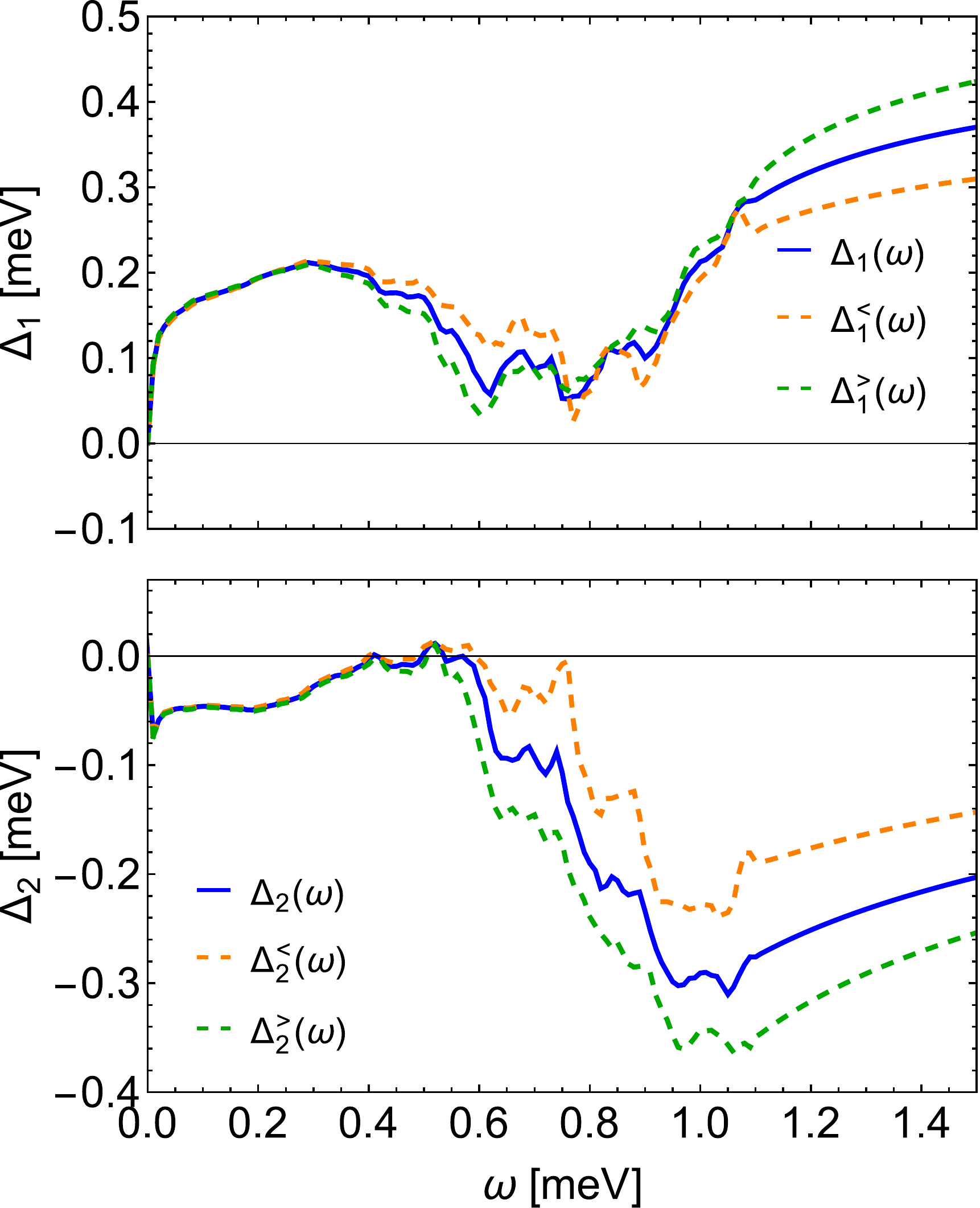}
\caption{Real and imaginary parts of the gap functions extracted using Eq.~\eqref{eq:gap_function} from the dos-functions plotted in Fig.~\ref{fig:3_dos_functions}. The same color coding has been used as in Fig.~\ref{fig:3_dos_functions}. Only the $\omega>0$ part of the functions is shown.}
\label{fig:3_delta}
\end{figure}

{\it Direct extraction.} In Fig.~\ref{fig:3_delta} we show the results of the direct extraction of the gap functions from the complex dos-functions plotted in Fig.~\ref{fig:3_dos_functions}. Note that up to $\omega\approx 0.3$~meV the gap functions are essentially independent of the relative norm of $N_{se}(\omega)$ and $N_{ne}(\omega)$ for all acceptable values of this norm, but also at higher energies the qualitative features are the same in all solutions: a steep rise of $\Delta_1(\omega)$ and a negative peak of $\Delta_2(\omega)$, both taking place around $\omega=\omega_\ast$. These are precisely the features of the simple model gap function Eq.~\eqref{eq:model_delta}, as expected.

\begin{figure}[t]
\includegraphics[width = 7.5 cm]{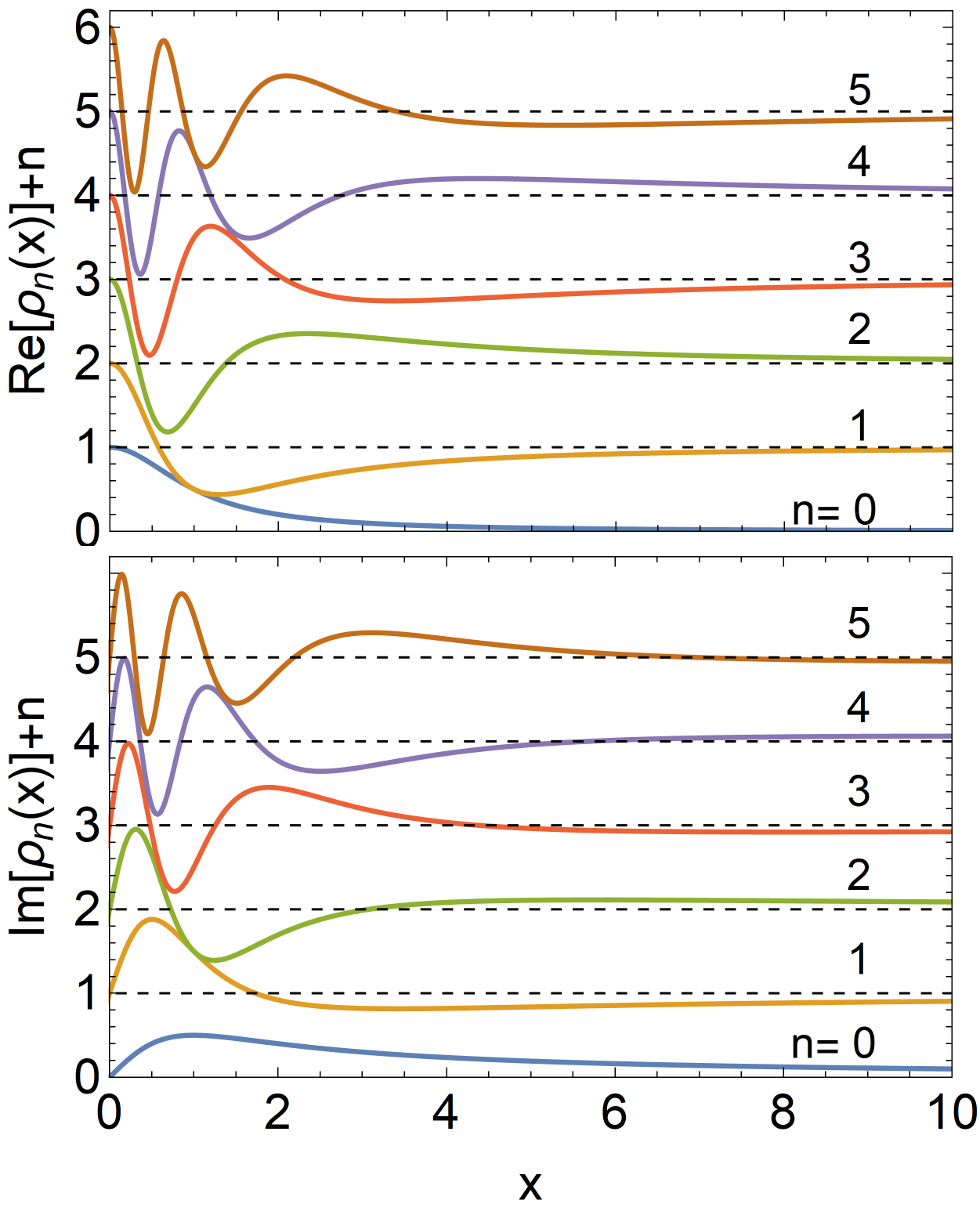}
\caption{Real (upper panel) and imaginary (lower panel) parts of the basis functions $\rho_n(x)$ for $n=0,\ldots,5$. Note that, with increasing $n$, the period of oscillations at small $x$ increases and, at the same time, the largest nodes $x_n$ of both components grow.}
\label{fig:basis_functions}
\end{figure}

{\it Extraction by expansion.} For the sake of completeness, in Fig.~\ref{fig:basis_functions} we plot the first six basis functions $\rho_n(x)$ of the complete orthogonal basis introduced in \cite{Weideman95}. Note that this basis is well suited for expansion of functions which have a complicated small-$x$ behavior, but at large values of $x$ they are relatively smooth. This has implications for the optimal choice of the energy scale $E$ in Eq.~\eqref{eq:gap_expansion}. In order to describe well the Dynes-like features close to the Fermi level, we have chosen to take $E=0.1$~meV.

\begin{figure}[b]
\includegraphics[width = 7.5 cm]{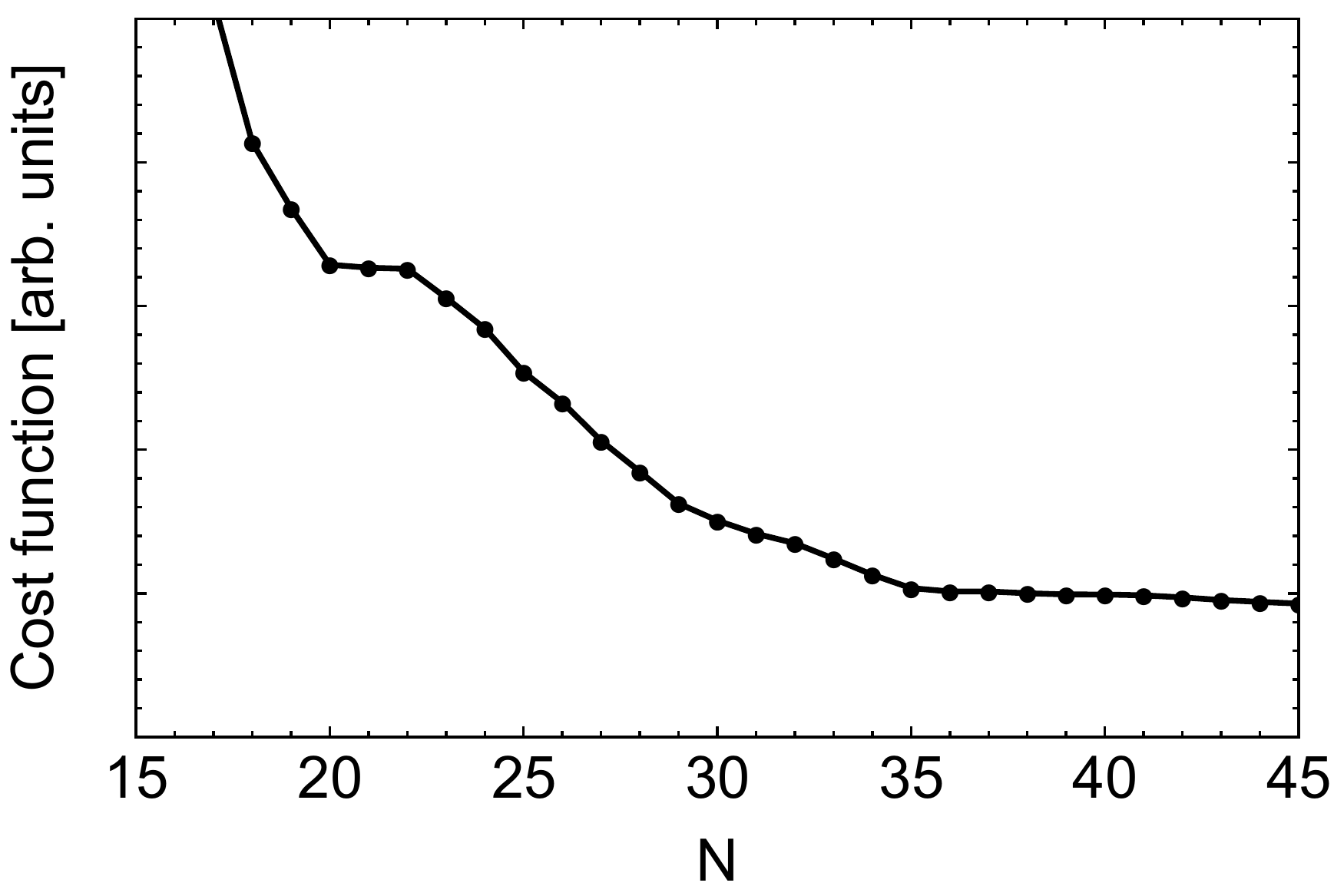}
\caption{Dependence of the cost function on the order $N$ of the expansion for $E=0.1$~meV.}
\label{fig:cost_function}
\end{figure}

In order to fix the optimal value of $N$ for this choice of $E$, we study how the cost function varies with $N$. The result of this calculation is shown in Fig.~\ref{fig:cost_function}. As expected, the cost function decreases monotonically with $N$. In the limit of large $N$ the cost function should vanish, because in that limit Eq.~\eqref{eq:gap_expansion} reproduces exactly the direct solution. 

What we are after is a solution with the least possible $N$ which does reproduce faithfully the measured complex dos-function. Based on Fig.~\ref{fig:cost_function}, we have chosen to take $N=35$, since the decay of the cost function for $N>35$ is slow, indicating that inclusion of further basis functions in the expansion Eq.~\eqref{eq:gap_expansion} improves only the agreement with noise-like features in $n(\omega)+i\kappa(\omega)$.

\begin{figure}[t]
\includegraphics[width = 7.5 cm]{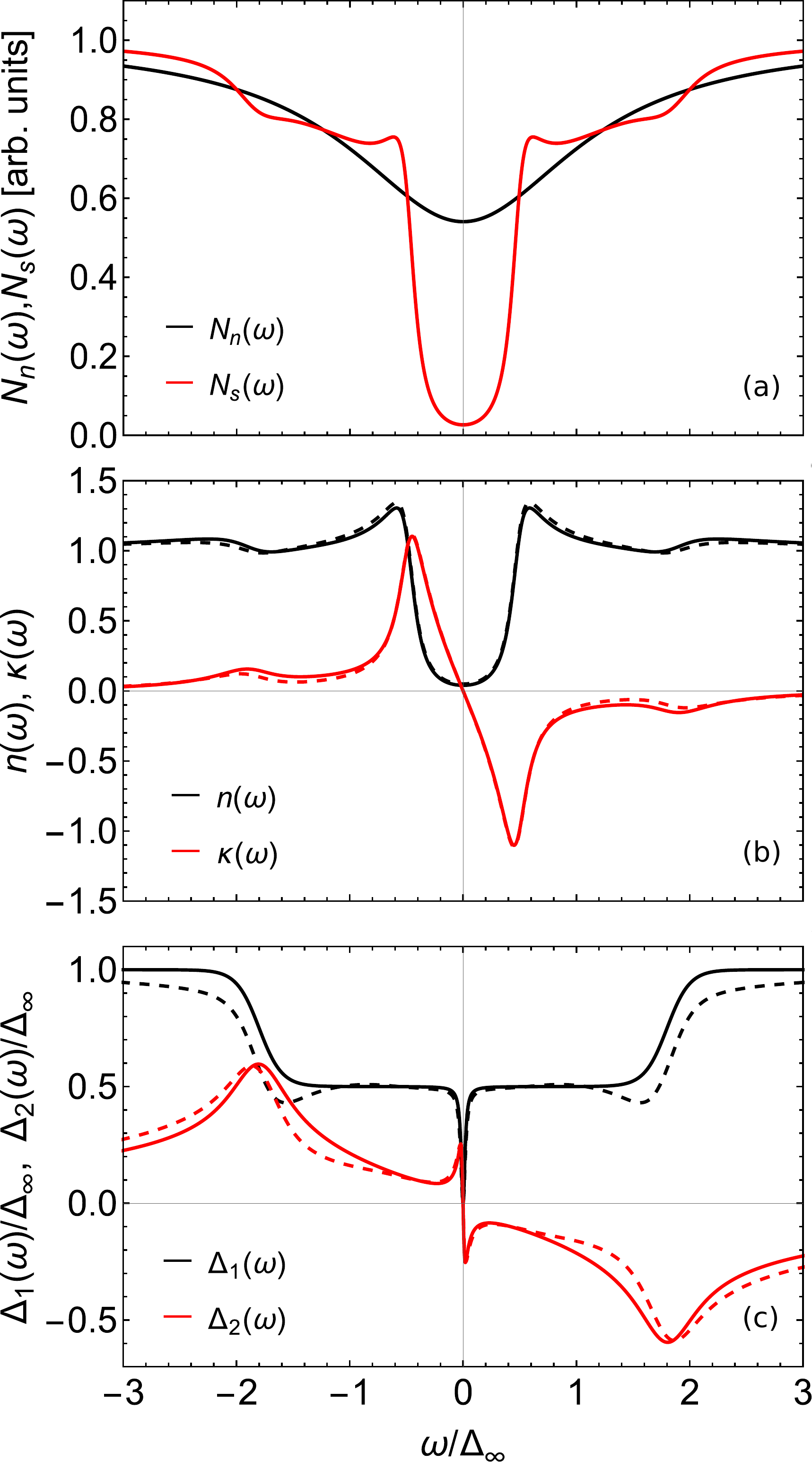}
\caption{Test of the inversion procedure using model data, see text for more details. The auxiliary function $N_0(\varepsilon)$ is parametrized by $\alpha/\Delta_{\infty}=1.76$ and $V/\Delta_{\infty}=1.2$. The gap function is parametrized by $\Delta_0/\Delta_{\infty}=0.5$, $\omega_\ast/\Delta_{\infty}=1.8$, $\Theta/\Delta_{\infty}=0.1$, and $\Gamma/\Delta_{\infty}=0.02$. (a): exact densities of states in the normal and superconducting states, $N_n(\omega)$ and $N_s(\omega)$. (b): exact (solid) and extracted (dashed) dos-functions $n(\omega)+i\kappa(\omega)$. (c): exact (solid) and extracted (dashed) gap functions $\Delta(\omega)$.}
\label{fig:test}
\end{figure}

\section{Test of the extraction procedure}
\label{sec:test}
Our analysis is based on Eq.~(\ref{eq:change}) which is valid only approximately. In order to check the quality of our procedure, finally we perform the following test with parameters chosen so as to resemble the experimental data. 

We consider a conductor with a suppressed density of states in the vicinity of the Fermi level, described by a model auxiliary function $N_0(\varepsilon)\propto 1-\alpha \delta_{V}(\varepsilon)$. The parameters $\alpha$ and $V$ measure the depth and the width of the suppression, respectively. 

For the gap function of the superconductor, we take
$$
\Delta(\omega)=\Delta_\infty + (\Delta_0-\Delta_\infty)F(\omega)
-\frac{i\Gamma \Delta_D}{\omega+i\Gamma}.
$$
The first two terms reproduce the gap function Eq.~\eqref{eq:model_delta} characterized by two gap values $\Delta_0$ and $\Delta_\infty$, with a crossover between the two characterized by its position $\omega_\ast$ and width $\Theta$. In order to produce results which are similar to the experimental ones, we take $\Delta_\infty\approx 0.456$~meV, as required by the prolongation parameter $a\approx 0.104$~meV$^2$. The last term, in which we take $\Delta_D=\Delta_\infty+(\Delta_0-\Delta_\infty)F(0)\approx \Delta_0$, generates at low energy a Dynes-like feature in $\Delta(\omega)$ characterized by the parameter $\Gamma$. Also for the wave-function renormalization we take a Dynes-like expression $Z(\omega)=1+i\Gamma/\omega$.

Once the functions $\Delta(\omega)$ and $Z(\omega)$ are known, we can calculate the spectral functions via Eq.~\eqref{eq:momentum}. The densities of states in the normal and superconducting states of the test model, $N_n(\omega)$ and $N_s(\omega)$, can be found by numerically taking the integral in Eq.~\eqref{eq:dos_definition}, since the auxiliary function $N_0(\varepsilon)$ is also known.  The results are shown in panel (a) of Fig.~\ref{fig:test}. Note the qualitative resemblance to experimental data shown in Fig.~\ref{fig:normalization}.

Next, applying the numerical inversion procedure based on Eqs.~(\ref{eq:change},\ref{eq:dos}) and taking $N_n(\omega)$ and $N_s(\omega)$ as input data, we determine the function $\Omega=\Omega(\omega)$ and, consequently, also the complex dos-function $n(\omega)+i\kappa(\omega)$. In panel (b) of Fig.~\ref{fig:test} we compare the obtained result with the exact dos-function calculated directly from the gap function $\Delta(\omega)$ using Eq.~\eqref{eq:dos}. Note that the agreement of our procedure with the exact data is very good. 

Finally, from the complex dos-function we extract the gap function using Eq.~\eqref{eq:gap_function}. In panel (c) of Fig.~\ref{fig:test}, the result of this extraction is compared with the original gap function $\Delta(\omega)$. One can observe that, while the agreement is not perfect, all features of the gap function are reproduced correctly. This completes our test of Eqs.~(\ref{eq:change},\ref{eq:dos}).


\begin{thebibliography}{99}
\bibitem{note_mott}This statement does not apply at strong coupling when the interaction strength exceeds the bandwidth.
\bibitem{note_born}An explicit example where the self-consistent Born approximation is insufficient is provided by the Dynes superconductors \cite{Kavicky20}.
\end{thebibliography}

\end{document}